\documentclass{article}
\usepackage{amsmath,graphicx,url,times,booktabs, tabularx}
\usepackage{authblk}

\title{HODGEPODGE: Sound event detection based on ensemble of semi-supervised learning methods}

\author[1]{Ziqiang Shi\thanks{Corresponding author: shiziqiang@cn.fujitsu.com; shiziqiang7@gmail.com}}
\author[1]{Liu Liu}
\author[1]{Huibin Lin}
\author[1]{Rujie Liu}
\author[2]{Anyan Shi}

\affil[1]{Fujitsu Research and Development Center, Beijing, China}
\affil[2]{Shuangfeng First, Beijing, China}

\begin{document}

\maketitle

\begin{sloppy}

\begin{abstract}
In this paper, we present a method called HODGEPODGE\footnotemark[1] for large-scale detection of sound events using
weakly labeled, synthetic, and unlabeled data proposed in the Detection and Classification of Acoustic Scenes and Events
(DCASE) 2019 challenge Task 4: Sound event detection
in domestic environments. To perform this task, we
adopted the convolutional recurrent neural networks (CRNN) as our backbone network.
In order to deal with a small amount of tagged data and a large amounts of unlabeled in-domain data,
we aim to focus primarily on how to apply semi-supervise learning methods efficiently to make full use of limited data.
Three semi-supervised learning principles have been used in our system, including: 1)
Consistency regularization applies data augmentation; 2) MixUp regularizer requiring
that the predictions for a interpolation of two inputs is close to the interpolation
of the prediction for each individual input; 3) MixUp regularization applies to
interpolation between data augmentations.
We also tried an ensemble of
various models, which are trained by using different semi-supervised learning principles.
Our proposed approach significantly improved
the performance of the baseline, achieving the event-based
f-measure of 42.0\% compared to 25.8\% event-based f-measure of
the baseline in the  provided official evaluation dataset.
Our submissions ranked third among 18 teams
in the task 4.
\end{abstract}
\footnotetext[1]{HODGEPODGE has two layers of meanings. The first layer is the variety of training data involved in the method, including weakly labeled, synthetic, and unlabeled data. The second layer refers to several semi-supervised principles involved in our method.}

\section{Introduction}
\label{sec:intro}

The sound carries a lot of information about our everyday environment and the physical events that take place there. We can easily perceive the sound scenes we are in (busy streets, offices, etc.) and identify individual sound events (cars, footsteps, etc.). The automatic detection of these sound events has many applications in real life. For example, it's very useful for intelligent devices, robots, etc., in the environment awareness. Also a sound event detection system can help to construct a complete monitoring system when the radar or video system may not work in some cases.

To contribute to the sound event detection task,
the Detection and
Classification of Acoustic Scenes and Events (DCASE) challenge has been organized for four
years since 2013~\cite{mesaros2018detection,mesaros2017dcase,dcase2019web}.
DCASE is a series of
challenges aimed at developing sound classification and detection
systems~\cite{mesaros2018detection,mesaros2017dcase,dcase2019web}.
This year, the DCASE 2019 challenge
comprises five tasks: acoustic scene classification, audio tagging with noisy labels and minimal supervision,
sound event localization and detection, sound event detection in domestic environments, and urban sound tagging~\cite{dcase2019web}.
Among them, this paper describes a method for performing the task 4 of the DCASE
2019 challenge, large-scale detection of sound events in domestic environments using real data either weakly labeled or unlabeled, and synthetic data that is strongly labeled (with time stamps). The aim is to predict
the presence or absence and the onset and offset times of sound
events in domestic environments. This task is the follow-up to DCASE 2018 task 4, which
aims at exploring the possibility to exploit a large amount of unbalanced and unlabeled training data together with a small weakly annotated training set to improve system performance. The difference is that there is an additional training set with strongly annotated synthetic data is provided in this year's task 4. Thus it can be seen that we are faced with three difficult problems: 1) there is no real strongly labeled and only too few weakly labeled data, 2) the synthetic data is obviously different from the real one, and how is the effect of synthetic data on the detection results? and 3) there is too much unlabeled data.
Although this task is difficult , there have been a variety of methods proposed to solve this problem~\cite{serizel2018large,jiakai2018mean,kong2018dcase}.
Furthermore, a baseline
system that performs the task is provided in the DCASE 2019
challenge~\cite{tarvainen2017mean,jiakai2018mean}.

Based on these previous studies, we propose to apply a convolutional
recurrent neural network (CRNN), which is used as the backbone network in the baseline system
for task 4 of DCASE 2019~\cite{dcase2019web}. In order to make full use of small amount of weakly labeled and synthetic data, the principles in interpolation consistency training (ICT)~\cite{verma2019interpolation} and MixMatch~\cite{berthelot2019mixmatch} has been adopted in the `Mean Teacher'~\cite{tarvainen2017mean, jiakai2018mean} framework.
To avoid overfitting, consistency regularization on the provided unlabeled data is incorporated.

The rest of this paper is organized as follows:  Section~\ref{sec:method} introduces details of our proposed HODGEPODGE. The experiment
settings and results are displayed and discussed in Section~\ref{sec:results}. We conclude this
paper in Section~\ref{sec:conc}.

\section{Proposed method}
\label{sec:method}

Herein, we present the method of our submissions for task 4 of DCASE 2019. In the following sections, we will describe the details of our approach, including feature extraction, network structure, how to use ICT and MixMatch in the context of `Mean Teacher', and how to use unlabeled data.

\subsection{Feature extraction}
\label{ssec:preprocessing}

The dataset for task 4 is composed of 10 sec audio clips recorded in domestic environment
or synthesized to simulate a domestic environment.
No preprocessing step was applied in the presented frameworks.
The acoustic features for the 44.1kHz original data used in this system consist of 128-dimensional log mel-band energy extracted in Hanning windows of size 2048 with 431 points overlap. Thus the maximum number of frames is 1024.
In order to prevent the system from overfitting on the small amount of development data, we added random white noise (before log operation) to the melspectrogram  in each mini-batch during training. The input to the network is fixed to be 10-second audio clip. If the input audio is less than 10 seconds, it is padded to 10 seconds; otherwise it is truncated to 10 seconds.

\subsection{Neural network architecture}
\label{ssec:architecture}

\begin{figure*}[htb]
\centering
\includegraphics[width=0.9\textwidth]{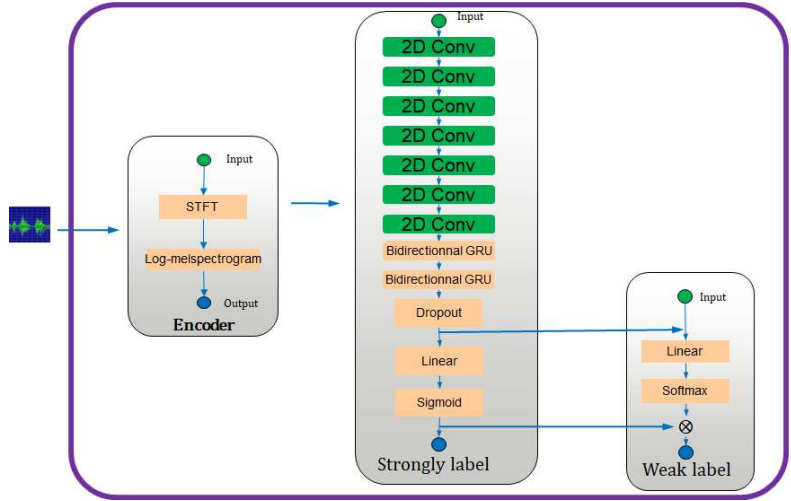}
\caption{Architecture of the CRNN in HODGEPODGE.}
\label{fig:crnn}
\end{figure*}

Figure~\ref{fig:crnn} presents the CRNN network architecture
employed in our HODGEPODGE. The audio signal is first converted to [128$\times$1024] log-melspectrogram to form the input to the network.
The first half of the network consists of the seven convolutional layers, where we use gated linear units (GLUs) instead of commonly rectified linear units (RELUs) or leaky ReLUs as nonlinear activations.

\begin{figure}[htb]
\centering
\includegraphics[width=0.45\textwidth]{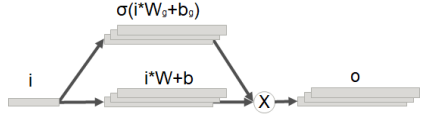}
\caption{Architecture of a GLU.}
\label{fig:gcnns}
\end{figure}

Figure~\ref{fig:gcnns} shows the structure of a GLU :
\begin{equation}
o=(i*W+b)\otimes \sigma(i*W_g+b_g), \nonumber
\label{eq:gcnns}
\end{equation}
where $i$ and $o$ are the input and output, $W$, $b$, $W_g$, and $b_g$ are learned parameters, $\sigma$ is the sigmoid function and $\otimes$ is the element-wise product between vectors or matrices.
Similar to LSTMs, GLUs play the role of controlling the information passed on
in the hierarchy.
This special gating mechanism allows us to effectively capture long-range context dependencies by deepening layers without encountering the problem of vanishing gradient.

For the seven gated convolutional layers, the kernel sizes are 3, the paddings are 1, the strides are 1, and the number of filters are [16, 32, 64, 128, 128, 128, 128] respectively, and the poolings are [(2, 2), (2, 2), (1, 2), (1, 2), (1, 2), (1, 2), (1, 2)] respectively. Pooling along the time axis is used in
training with the clip-level and frame-level labels.

The gated convolutional blocks are followed by two bidirectional gated recurrent units (GRU) layers
containing 64 units in the forward and backward path, their output
is concatenated and passed to the attention and classification layer
which are described below.

As depicted in Figure~\ref{fig:crnn}, the output of the bidirectional GRU layers is fed
into both a frame-level classification block and an attention block respectively. The frame-level classification
block uses a sigmoid activation function to predict the probability of
each occurring class at each frame. Thus bidirectional GRUs followed by a dense
layer with sigmoid activation to compute posterior probabilities of
the different sounds classes.
In that case there are two outputs in this CRNN.
The output from bidirectional GRUs followed by dense layers with sigmoid activation
is considered as sound event detection result.  This output can be
used to predict event activity probabilities. The other output is the
weighted average of the element-wise multiplication
of the attention, considering as audio
tagging result.
Thus the final prediction for the weak label of each class is determined
by the weighted average of the element-wise multiplication
of the attention and classification block output of each class c.

\subsection{Semi-supervised learning}
\label{ssec:ssl}

Inspired by the DCASE 2018 task 4 winner solution~\cite{jiakai2018mean} and the baseline system~\cite{turpault2019sound}, in which it uses the `Mean Teacher' model~\cite{tarvainen2017mean}. `Mean Teacher' is a combination of two models: the student model and the teacher model.
At each training step, the student model is trained on synthetic and weakly labeled data with binary cross entropy classification cost.
While the teacher model uses the exponential moving average of the student model. The student model is the final model and the teacher model is designed to help the student model by a consistency mean-squared error cost
for frame-level and clip-level predictions of unlabeled audio clips. That means good student should output the same class distributions as the teacher for the same unlabeled example even after it has been augmented.
The goal of `Mean Teacher' is to minimize:
\begin{equation}
L=L_w+L_s+w(t)L_{cw}+w(t)L_{cs} \nonumber
\end{equation}
where $L_w$ and $L_s$ are the usual cross-entropy classification loss on weakly labeled data with only weak labels and synthetic data with only strong labels respectively, $L_{cw}$ and $L_{cs}$ are the teacher-student consistence regularization loss on unlabeled data with predicted weak and strong labels respectively, and $w(t)$ is the balance of classification loss and the consistency loss. Generally the $w(t)$ changes over time to make the  consistency loss initially accounts for a very small proportion, and then the ratio slowly becomes higher. Since in the beginning,  neither the student model nor the teacher model were accurate on predictions, and the consistency loss did not make much sense. $w(t)$ has a maximum upper bound, that is, the proportion of consistent loss does not tend to be extremely large. With different maximum upper bound of consistence weight $w(t)$, the trained model has different performances. n the next section, we ensemble the models trained under different maximum consistence weights to achieve better results.

HODGEPODGE did not change the overall framework of the baseline. It only attempts to combine several of the latest semi-supervised learning methods under this framework.

The first attempt is the interpolation consistency training (ICT) principle~\cite{verma2019interpolation}. ICT learns a student
network in a semi-supervised manner. To this end, ICT uses a `Mean Teacher' $f_{\theta'}$. During training, the student parameters $\theta$ are updated to encourage consistent predictions
\begin{equation}
f_\theta(\text{Mix}_\lambda(u_j,u_k))\approx \text{Mix}_\lambda(f_{\theta'}(u_j),f_{\theta'}(u_k)), \nonumber
\end{equation}
and correct predictions for labeled examples, where
\begin{equation}
\text{Mix}_\lambda(a,b)=\lambda a + (1-\lambda)b \nonumber
\end{equation}
is called the interpolation or MixUp~\cite{zhang2017mixup}. \nonumber
In our system, we perform interpolation
of sample pair and their corresponding labels (or pseudo labels predicted by the CRNNs) in both the supervised loss on labeled examples and the consistency loss on unsupervised examples.
In each batch, the weakly labeled data, synthetic data, and unlabeled data are shuffled separately to form a new batch.
Then use the ICT principle to generate new augmented data and labels with the corresponding clips in the original and new batches.
It should be noted that the $\lambda$ is different for each batch. Thus the loss
\begin{equation}
L_{ict}=L_{w,ict}+L_{s,ict}+w(t)L_{cw,ict}+w(t)L_{cs,ict} \nonumber
\end{equation}
where $L_{w,ict}$ and $L_{s,ict}$ are the classification loss on weakly labeled data with only weak labels and synthetic data  with only strong labels using ICT respectively, $L_{cw,ict}$ and $L_{cs,ict}$ are the teacher-student consistence regularization loss on ICT applied on unlabeled data with predicted weak and strong labels respectively.

The second try draws on some of the ideas in MixMatch~\cite{berthelot2019mixmatch}, but not exactly the same. MixMatch introduces a single loss that unifies entropy minimization, consistency regularization, and generic regularization approaches to semi-supervised learning. Unfortunately MixMatch can only be used for one-hot labels, not suitable for task 4, where there may be several events in a single audio clip.  So we didn't use MixMatch in its original form. In each batch, $K(>1)$ different augmentations are generated, then the original MixMatch does mixup on all data, regardless of whether the data is weakly labeled, synthetic or unlabeled. But our experiment found that the effect is not good, so we fine-tuned the MixMatch to do MixUp only between the augmentations of the same data type. The loss function is similar to the loss in the ICT case.


\subsection{Model ensemble and submission}
\label{ssec:ensemble}

To further improve the performance of the system, we use some ensemble methods to fuse different models.
The main differences of the single models have two dimensions, one is the difference of the semi-supervised learning method, and the other is the difference of the maximum value of the consistency loss weight.
For this challenge, we submitted 4 prediction results with different model ensemble:
\begin{itemize}
\item
HODGEPODGE 1: Ensemble model is conducted
by averaging the outputs of 9 different models with different maximum consistency coefficients in `Mean Teacher' principle. The F-score  on validation data was 0.367. (Corresponding to Shi\_BossLee\_task4\_1 in official submissions)
\item
HODGEPODGE 2 : Ensemble model is conducted
by averaging the outputs of 9 different models with different maximum consistency coefficients in ICT principle. The F-score  on validation data was 0.425. (Corresponding to Shi\_BossLee\_task4\_2 in official submissions)
\item
HODGEPODGE 3: Ensemble model is conducted
by averaging the outputs of 6 different models with different maximum consistency coefficients in MixMatch principle. The F-score  on validation data was 0.389. (Corresponding to Shi\_BossLee\_task4\_3 in official submissions)
\item
HODGEPODGE 4: Ensemble model is conducted
by averaging the outputs of all the 24 models in Submission 1, 2, and 3. The F-score  on validation data was 0.417. (Corresponding to Shi\_BossLee\_task4\_4 in official submissions)
\end{itemize}

\section{EXPERIMENTS AND RESULTS}
\label{sec:results}

\subsection{Dataset}

Sound event detection in domestic environments [11] is a task to
detect the onset and offset time steps of sound events in domestic
environments. The datasets are from  AudioSet~\cite{gemmeke2017audio}, FSD~\cite{fonseca2017freesound} and
SINS dataset~\cite{dekkers2017sins}]. The aim of this task is to investigate whether
real but weakly annotated data or synthetic data is sufficient for
designing sound event detection systems.
There are a total of
1578 real audio clips with weak labels,
2045 synthetic audio clips with strong labels, and 14412 unlabeled in domain audio clips in the development set, while the evaluation set contains 1168 audio clips.
Audio recordings
are 10 seconds in duration and consist of polyphonic sound events
from 10 sound classes.

\subsection{Evaluation Metric}

The evaluation metric for this task is based on the event-based F-score~\cite{mesaros2016metrics}. The predicted events are compared to a list of reference events by comparing the onset and offset of the predicted event to the overlapping reference event. If the onset of the predicted event is within 200 ms collar of the onset of the reference event and its offset is within 200 ms or 20\% of the event length collar around the reference offset, then the predicted event is considered to be correctly detected, referred to as true positive. If a reference event has no matching predicted event, then it is considered a false negative. If the predicted event does not match any of the reference events, it is considered a false positive. In addition, if the system partially predicts an event without accurately detecting its onset and offset, it will be penalized twice as a false positive and a false negative. The following equation shows the calculation of the F-score for each class.
\begin{equation}
F_c= \frac{2TP_c}{2TP_c+FP_c+FN_c}, \nonumber
\end{equation}
where $F_c, TP_c, FP_c, FN_c$ are the F-score, true positives, false
positives, false negatives of the class c respectively. The final evaluation
metric is the average of the F-score for all the classes.

\subsection{Results}

First we did some experiments to determine the best size of the median window. The median window is used in the post-processing of posterior probabilities to results in the final events with onset and offset. Table~\ref{tab:median_window} shows the performance of HODGEPODGE
systems on validation data set under different median window size. Coincidentally, all methods achieve the best performance when the window size is 9.

Table~\ref{tab:fscore} shows the final macro-averaged event-based evaluation
results on the test set compared to the baseline system. In fact, HODGEPODGE 1 is the ensemble of baselines, the only difference is that we use a deeper network, as well as higher sampling rate and larger features. It can be seen that both ICT and MixMatch principles can improve performance, especially ICT, which performs best in all HODGEPODGE systems.

\begin{table}[th]
\caption[medianwindow]{The performance of HODGEPODGE
systems on validation data set under different median window size.}\label{tab:median_window}
\centering
\begin{tabular}{|c|c|c|c|c|c|}
\hline
Median window size  & 5 & 7  & 9 & 11  & 13\\
\hline
HODGEPODGE 1 & 35.7\%	&36.4\% & 36.7\%	&36.5\%&36.1\%\\
\hline
HODGEPODGE 2 & 41.4\%	 &42.1\% & 42.5\%	&42.2\% &42.1\%\\
\hline
HODGEPODGE 3 & 38.1\%	 &38.7\% & 38.9\%	&38.3\% &37.9\%\\
\hline
HODGEPODGE 4 & 40.8\%	 &41.5\% & 41.7\% &41.3\% &40.9\%\\
\hline
\end{tabular}
\end{table}

\begin{table}[th]
\caption[fscore]{The performance of our approach compared to the baseline
system.}\label{tab:fscore}
\centering
\begin{tabular}{|c|c|c|}
\hline
Method & Evaluation & Validation \\
\hline
HODGEPODGE 1 & 37.0\%	&36.7\%\\
\hline
HODGEPODGE 2 & 42.0\%	 &42.5\%\\
\hline
HODGEPODGE 3 & 40.9\%	 &38.9\%\\
\hline
HODGEPODGE 4 & 41.5\%	 &41.7\%\\
\hline
Baseline & 25.8\%	&23.7\%  \\
\hline

\end{tabular}
\end{table}

\section{CONCLUSIONS}
\label{sec:conc}

In this paper, we proposed a method called HODGEPODGE for sound event detection using
only weakly labeled, synthetic and unlabeled data. Our approach is
based on CRNNs, whereby we introduce several latest semi-supervised learning methods, such as
interpolation consistence training and MixMatch into the `Mean Teacher' framework to leverage
the information in audio data that are not accurately labeled. The final F-score of our system on the evaluation set
is 42.0\%, which is significantly higher than the score of the
baseline system which is 25.8\%.

\bibliographystyle{IEEEtran}
\bibliography{refs}
%
%
%
%
%
%
%
%
%

\end{sloppy}
\end{document}